\documentstyle[12pt,epsfig]{article}
\begin{document}
 
\title{ In-medium properties of antikaons at finite temperature:
the $K^-/K^+$ ratio at GSI}

\author{L. Tol\'os$^1$, J.Schaffner-Bielich$^1$, A. Polls$^2$, A. Ramos$^2$ \\ \\
$^1$Institut f\"ur Theoretische Physik,
  J. W. Goethe-Universit\"at\\
  D-60054 Frankfurt am Main, Germany\\
$^2$Departament d'Estructura i Constituents de la Mat\`eria, \\
  Universitat de Barcelona, Diagonal 647, 08028 Barcelona, Spain}



\date{}

\maketitle

\vspace{0.5cm}
\begin{abstract}
  The $K^-/K^+$ ratio in heavy-ion collisions at GSI is
  studied including the properties of the participating hadrons in hot
  and dense matter, paying a particular attention to the in-medium properties of antikaons at finite temperature.
 The determination of the temperature and chemical
  potential at freeze-out conditions compatible with the ratio
  $K^-/K^+$ is very delicate, and depends on the approach adopted for
  the antikaon self-energy. Different approaches for the $K^-$ self-energy
 have been considered: non-interacting $K^-$, on-shell self-energy and
  single-particle spectral density.  We observe that
   the use of an energy dependent $\bar{K}$ spectral density
  lowers considerably the freeze-out temperature with respect to the on-shell approach. We also conclude that the full off-shell approach
  gives rise to the ``broad-band equilibration" advocated by Brown,
  Rho and Song.
\end{abstract}
\vspace{0.5cm}



\section{Introduction}

The study of the properties of hadrons in  hot and dense matter is receiving a lot of attention over the last years. A special effort has been invested to understand the properties of antikaons in the medium, especially after the speculation of the possible existence of an antikaon condensed phase in neutron stars \cite{Kaplan}.

Heavy-ion collisions at energies around 1-2 AGeV offer the possibility
of studying hadrons under extreme conditions \cite{Oeschler}.  
In particular, production and propagation of kaons and antikaons 
have been investigated with the kaon 
spectrometer (KaoS) of the SIS heavy-ion synchroton at GSI (Darmstadt).

One first observation in C$+$C and Ni$+$Ni collisions
\cite{Barth,Menzel} is that, as a function of the energy difference
$\sqrt{s}$-$\sqrt{s_{th}}$, where $\sqrt{s_{th}}$ is the energy for
the particle production (2.5 GeV for $K^+$ via $pp \rightarrow \Lambda
K^+ p$ and 2.9 GeV for $K^-$ via $pp \rightarrow p p K^-K^+$), the
number of $K^-$ balanced the number of $K^+$ although in $pp$
collisions the production cross-sections close to threshold are 2-3
orders of magnitude different. This could be interpreted as an
indication of an attractive $K^-$ optical potential in the medium, although a complementary explanation in terms of in-medium enhanced $\pi \Sigma \rightarrow K^- p$ production has also been suggested \cite{Juergi}.

It has also been observed that the $K^-/K^+$ ratio remains almost
constant for C$+$C, Ni$+$Ni and Au$+$Au collisions for 1.5 AGeV
\cite{Barth,Menzel,Forster}. This could indicate that the absorption
of $K^-$ via $K^-N \rightarrow Y \pi$ is suppressed and/or an enhanced
$K^-$ production is obtained because of an attractive $K^-$ optical
potential.

Finally, a centrality independence for the $K^-/K^+$ ratio has been
noticed in Au$+$Au and Pb$+$Pb reactions at 1.5 AGeV \cite{Menzel}. A
recent interpretation claims that this centrality independence is a
consequence of the strong correlation between the $K^+$ and $K^-$
yields \cite{Hartnack}.  In fact, the centrality independence of the
$K^-/K^+$ ratio has often been advocated as signalling the lack of
in-medium effects in the framework of statistical models as the volume
cancels out exactly in the ratio \cite{Cleymans}. However, Brown et
al. introduced the concept of ``broad-band equilibration''
\cite{Brown} according to which the independence on centrality of the
$K^-/K^+$ ratio can be explained including medium effects.

This paper is devoted to investigate the influence of dressing the
antikaons on the $K^-/K^+$ ratio with particular emphasis of bringing
new insight into the role of in-medium effects in heavy-ion collisions
at GSI energies.  

\section{In-medium effects on the \mbox{\boldmath$K^-/K^+$} ratio}

In this section we present a calculation of the $K^-/K^+$ ratio in the
framework of the statistical model. The basic hypothesis is to assume 
that the particle ratios in relativistic heavy-ion collisions can be described by two parameters, the baryonic chemical potential $\mu_B$ and the temperature $T$ \cite{Cleymans}.

The fact that the number of strange particles in the final state is
small at GSI energies requires an exact treatment of strangeness
conservation, while the baryonic and charge conservation laws can be
satisfied on average.

Therefore, using statistical mechanics, one obtains for the $K^-/K^+$
ratio \cite{Cleymans,Laura03}
\begin{eqnarray}
\frac{K^-}{K^+}=\frac{Z_{K^-}^1}{Z_{K^+}^1} \ \frac{Z_{K^+}^1+
Z_{M,S=+1}^1} {Z_{K^-}^1+  Z_{B,S=-1}^1+ Z_{M,S=-1}}
\label{eq:largevolume}
\end{eqnarray}
where $Z^1_{K^+}$ $(Z^1_{K^-})$ is the one-particle partition function for
$K^+(K^-)$, and $Z^1_{B,S=\pm 1}$ $(Z^1_{M,S=\pm 1})$ indicate the sum of
one-particle partition functions for baryons (mesons) with $S=\pm 1$.
It is interesting to note that the $K^-/K^+$ ratio is independent of the
volume and, hence, the same in the canonical and grandcanonical scheme
used for strangeness conservation in contrast to the $K^-$ and $K^+$
multiplicities alone \cite{Cleymans}.

Our objective is to study how the
in-medium modifications of the properties of the hadrons at finite
temperature affect the value of the $K^-$/$K^+$ ratio, focusing our
attention on the dressing of antikaons.  For consistency with previous
works, we prefer to compute the inverse ratio $K^+$/$K^-$
\begin{eqnarray}
\frac{K^+}{K^-}
=\frac{Z^1_{K^+}(Z^1_{K^-}+Z^1_{\Lambda}+
Z^1_{\Sigma}+Z^1_{\Sigma^*})}{Z^1_{K^-}Z^1_{K^+}}
=1+\frac{Z^1_{\Lambda}+Z^1_{\Sigma}+Z^1_{\Sigma^*}}{Z^1_{K^-}}
\label{eq:lamb-sig-k}
\end{eqnarray}
This expression is equivalent to Eq.~(\ref{eq:largevolume}) but
takes into account only the most relevant contributions. For balancing the
number of $K^+$, the main contribution in the $S=-1$ sector comes from
$\Lambda$ and $\Sigma$ hyperons and, in a smaller proportion, from
$K^-$ mesons and $\Sigma^*(1385)$ resonances.  The number of $K^-$ is
balanced by the presence of $K^+$ mesons.

Then, the particles involved in the calculation should be dressed
according to their properties in the hot and dense medium in which
they are embedded. For the $\Lambda$ and $\Sigma$, the partition
function
\begin{eqnarray}
Z_{\Lambda,\Sigma}=  g_{\Lambda,\Sigma}  V  \int
\frac{d^3p}{(2\pi)^3} e^{\frac{-\sqrt{m^2_{\Lambda,\Sigma}+p^2}-
U_{\Lambda,\Sigma}(\rho)+\mu_{B}}{T}} \ ,
\label{eq:lam-sig}
\end{eqnarray}
is built using a mean-field dispersion relation for the
single-particle energies (see 
Ref.~\cite{Balberg} for the dressing of $\Lambda$ and
Ref.~\cite{Mares} for $\Sigma$). The resonance
$\Sigma^*(1385)$ is described by a Breit-Wigner shape
\begin{eqnarray}
Z_{\Sigma*}&=& g_{\Sigma^*} V \int \frac{d^3p}{(2\pi)^3}
\int_{(m_{\Sigma^*}-2\Gamma)^2}^{(m_{\Sigma^*}+2\Gamma)^2}\ ds \
e^{\frac{-\sqrt{p^2+s}}{T}} \frac{1}{\pi}
\frac{m_{\Sigma^*}\Gamma}{(s-m_{\Sigma^*}^2)^2+m_{\Sigma^*}^2\Gamma^2}
\  e^{\frac{\mu_B}{T}}  ,
\label{eq:resonance}
\end{eqnarray}
with $m_{\Sigma^*}=1385 \ {\rm MeV}$ and $\Gamma=37 \ {\rm MeV}$.

\begin{figure}[htb]
  \centerline
  {\includegraphics[width=0.54\textwidth,angle=-90]{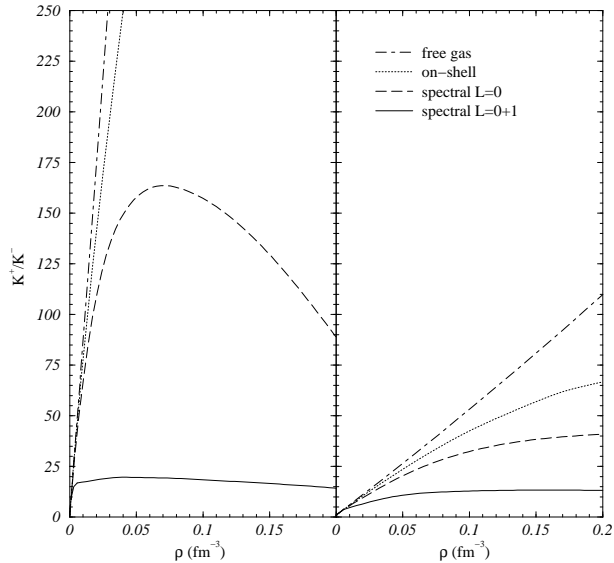} }
\caption{$K^+/K^-$ ratio as a function of density for $T=50$ MeV
  (left panel) and $T=80$ MeV (right panel) using different approaches
  to the $K^-$ optical potential: free gas (dot-dashed lines), the
on-shell approach (dotted lines) and using the $K^-$ spectral density
including s-waves (long-dashed lines) or both s- and p-waves (solid
lines).}
\label{fig:ratio-dens}
\end{figure}
Finally, two different prescriptions for the single-particle energy of
the antikaons have been used (see Fig.\ref{fig:antikaon}). First, we
use the on-shell or mean-field approximation for the $K^-$
potential in which the partition function reads
\begin{eqnarray}
Z_{K^-}&=& g_{K^-} V \int\frac{d^3p}{(2\pi)^3}
e^{\frac{-\sqrt{m_{K^-}^2 +p^2}-U_{K^-}(T,\rho,E_{K^-},p)}{T}} \ ,
\label{eq:onshell}
\end{eqnarray}
being $U_{K^-}(T,\rho,E_{K^-},p)$ the $K^-$ single-particle
potential in the Brueckner-Hartree-Fock approach
\cite{Laura02,Laura01} (see l.h.s. of Fig.~\ref{fig:antikaon}).
 The second approach incorporates the $K^-$ spectral
density,
\begin{eqnarray}
Z_{K^-}&=& g_{K^-} V \int \frac{d^3p}{(2\pi)^3} \int ds
\ S_{K^-}(p,\sqrt{s}) \ e^{\frac{-\sqrt{s}}{T}} \ , 
\label{eq:offshell}
\end{eqnarray}
using the s-wave component of the J\"ulich interaction and adding the
p-wave contributions as done in Ref.~\cite{Laura03} (see r.h.s. of
Fig.~\ref{fig:antikaon}). This p-wave components come from the
coupling of the $K^-$ meson to hyperon-hole states ($YN^{-1}$).

\section{Results for the \mbox{\boldmath $K^-/K^+$} ratio}

In this section we discuss the effects of dressing the $K^-$ mesons in
hot and dense matter on the $K^-/K^+$ ratio using an experimental value
of 0.031 $\pm$ 0.005 as reported in \cite{Menzel} for Ni$+$Ni
collisions at 1.93 AGeV.

\begin{figure}[htb]
\begin{minipage}
  {140mm}
  \includegraphics[height=0.53\textwidth,width=0.495\textwidth]{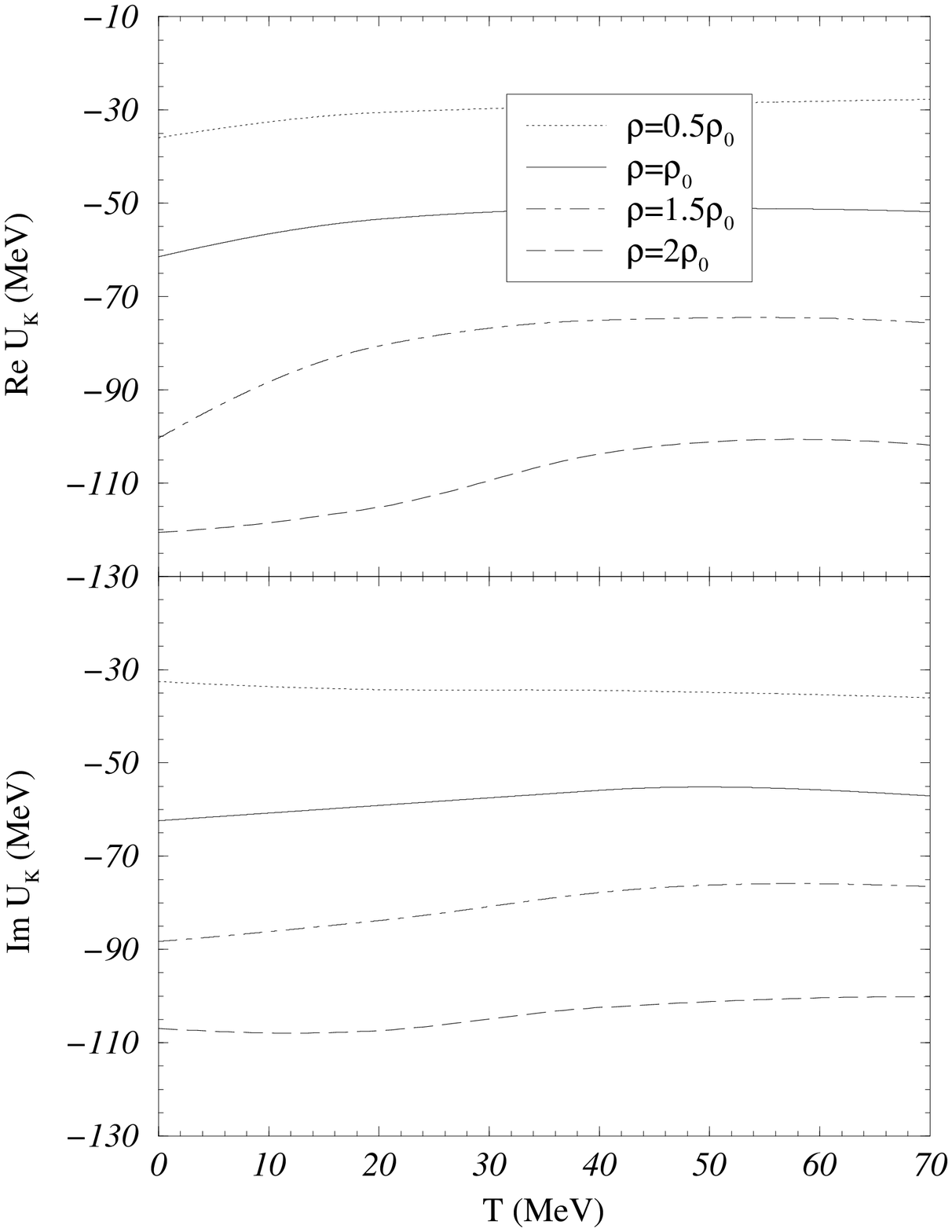}
  \includegraphics[height=0.53\textwidth,width=0.495\textwidth]{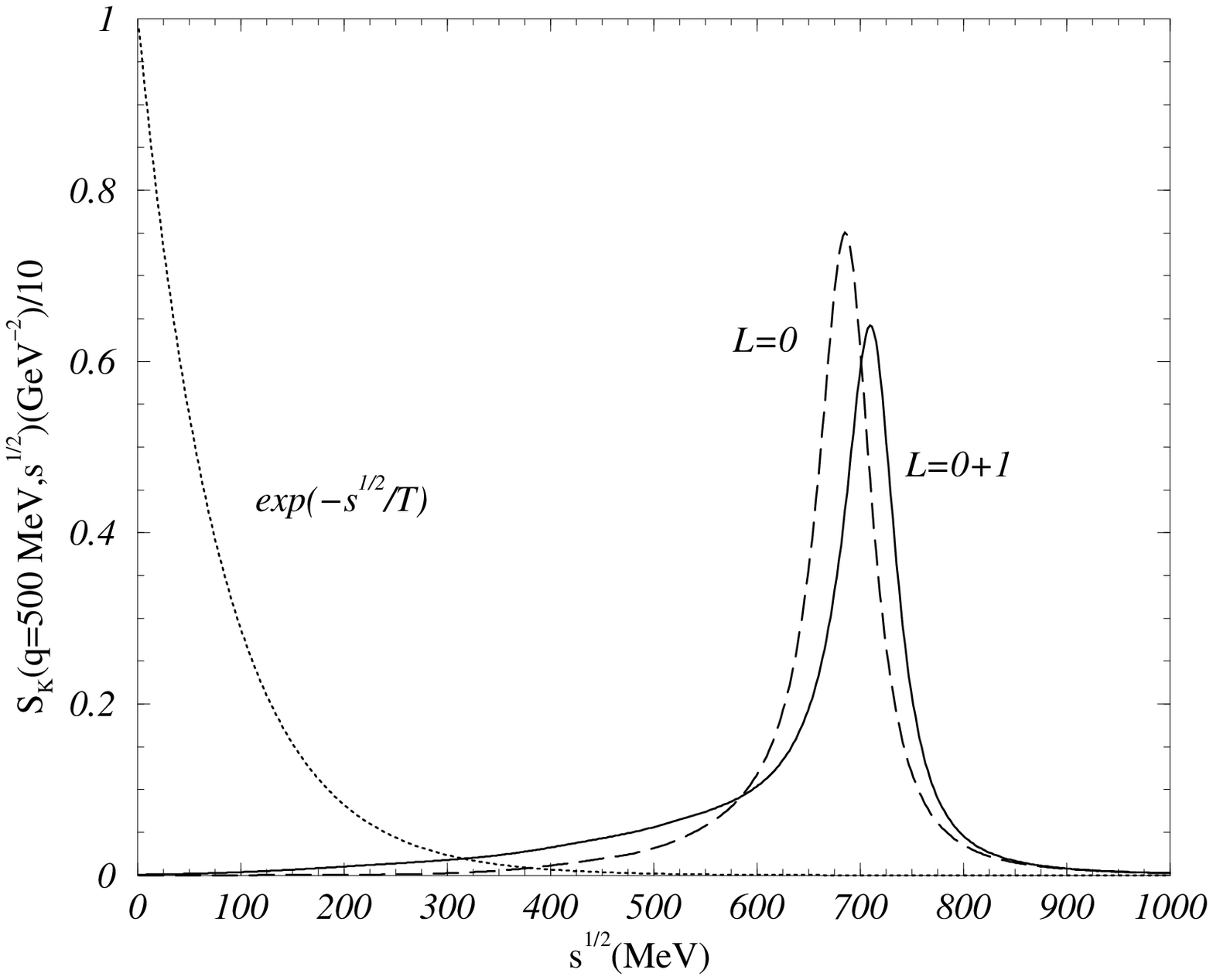}
\caption{Left: The real and imaginary part of the 
$K^-$ optical potential at zero momentum 
  for different densities as a function of temperature.  Right: The
  Boltzmann factor (dotted line) and the $K^-$ spectral function,
  including s-wave (dashed line) or s- and p-wave (solid line)
  components of the ${\bar K}N$ interaction, as functions of the
  energy, for a momentum $q=500$ MeV at saturation density and
  temperature $T= 80$ MeV. }
\label{fig:antikaon}
\end{minipage}
\end{figure}

The inverse ratio, $K^+/K^-$,
 is shown in Fig.~\ref{fig:ratio-dens} as a function of density at 
two given temperatures ($T=50$ MeV and $T=80$ MeV) using different approaches
for the dressing of the $K^-$ meson: free gas (dot-dashed lines), the
on-shell approach (dotted lines) and using the $K^-$ spectral density
including s-waves (long-dashed lines) or both s- and p-waves (solid
lines). Since the baryonic chemical potential $\mu_B$ grows with density, we note that the ratio grows with $e^{\mu_B/T}$ with increasing density.  The curves representing the $K^+/K^-$ ratio tend to bend
down after the initial increase when the in-medium $K^-$ properties are
included.  This effect is particularly notorious when the s- and p-wave
contributions of the $K^-$ self-energy are taken into account
in the spectral density. Actually, the low energy components of the
$K^-$ spectral density related to $YN^{-1}$ excitations are responsible
for this behaviour (see the overlap of the Boltzmann factor with the low
energy region of the $K^-$ spectral density in Fig.~\ref{fig:antikaon}).
These results are in qualitative agreement with the ``broad-band equilibration'' notion introduced by Brown et al.\cite{Brown}.  However, this behaviour was found using a
mean-field model, through a compensation of the increased
attraction of the mean-field $K^-$ potential with the increase in the
baryon chemical potential as density grows. In contrast, our mean-field
approach does not achieve this ``broad-band'' behaviour.

\begin{figure}[htb]
\begin{minipage}
  {140mm} \includegraphics[height=0.55\textwidth,width=0.495\textwidth]{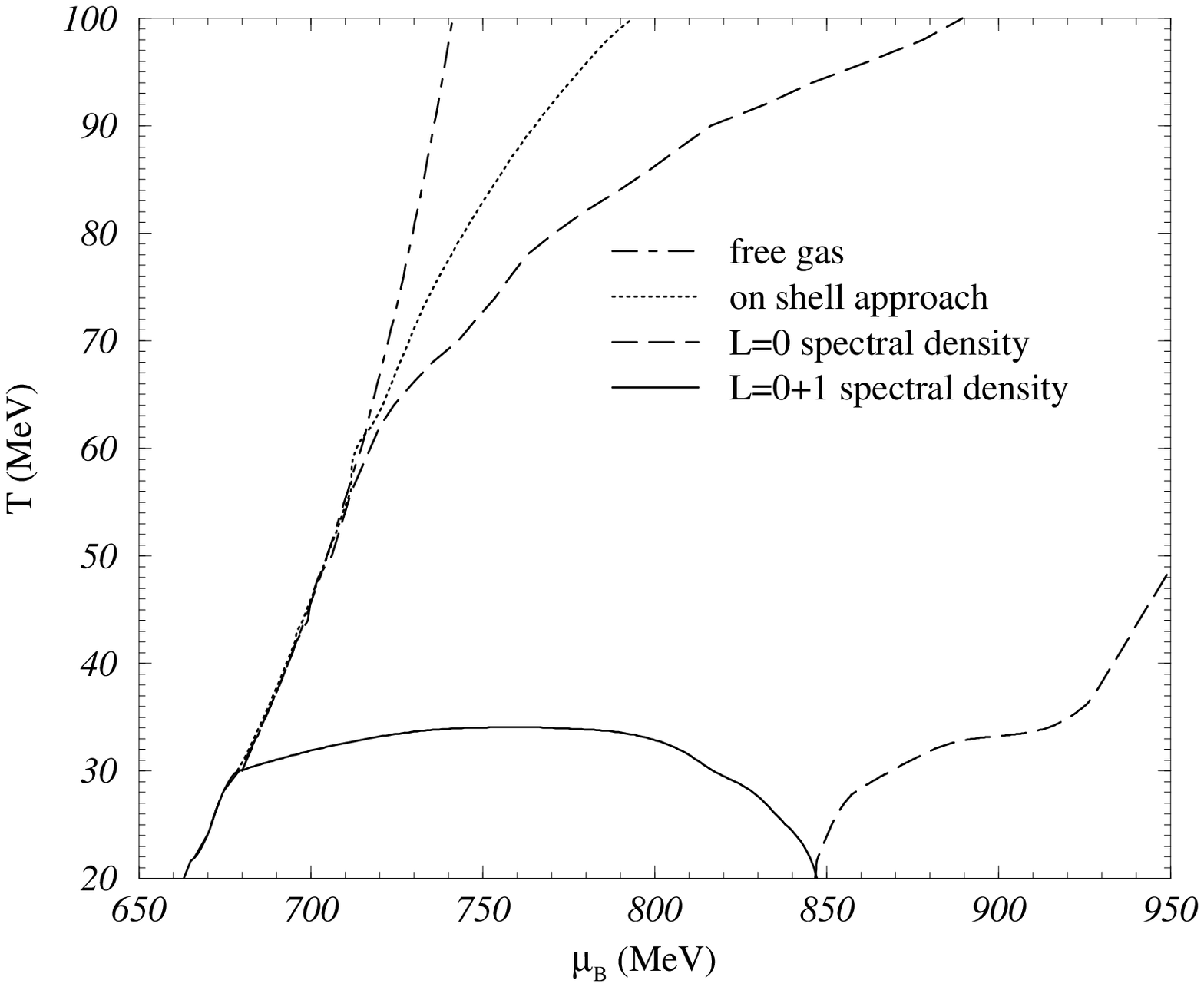}
  \includegraphics[height=0.55\textwidth,width=0.495\textwidth]{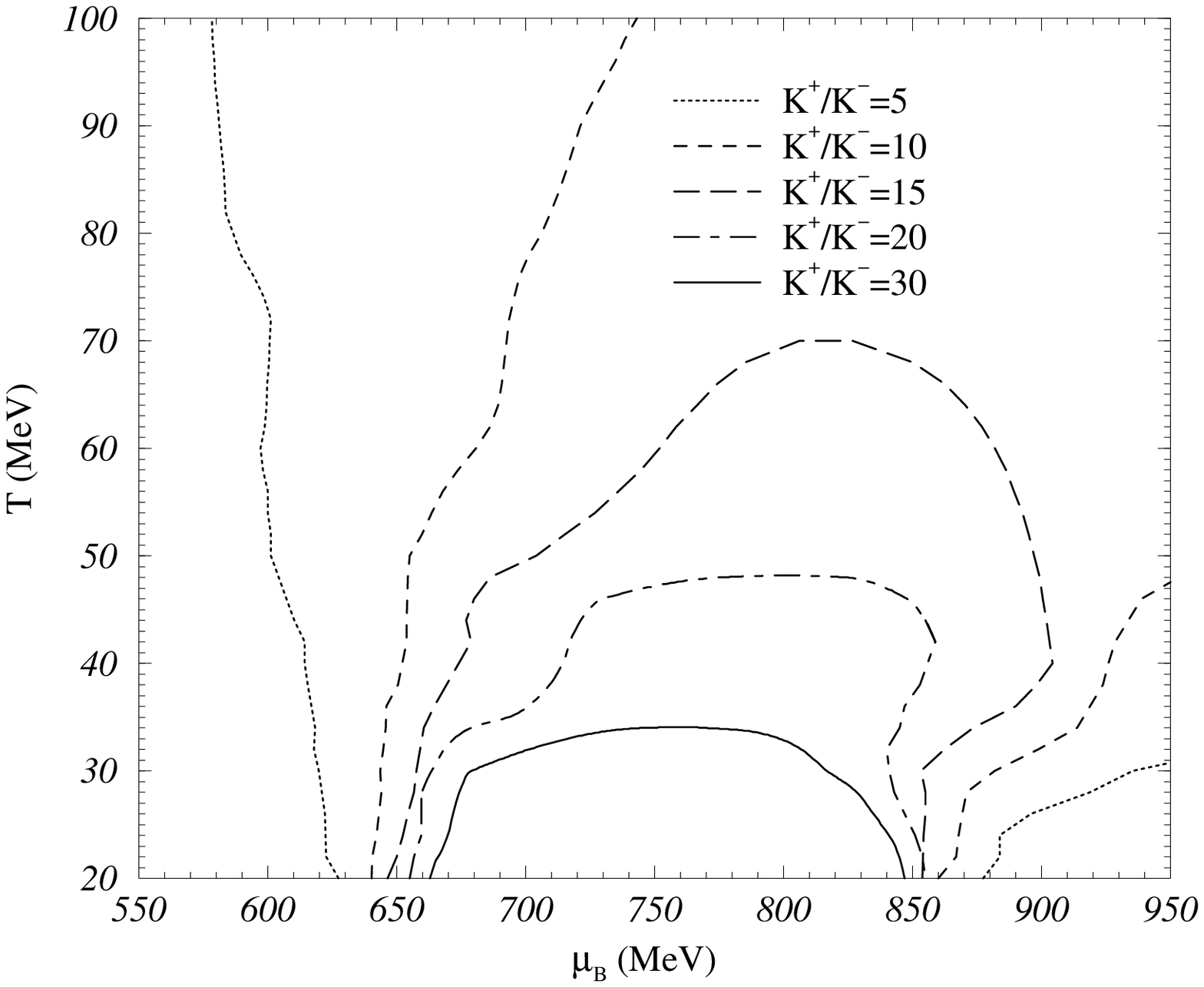}

\caption{Left: $T(\mu_B)$ for $K^+/K^-=30$ within different approaches. 
  Right: $T(\mu_B)$ for different values of the $K^+/K^-$ ratio
 using the full $K^-$ spectral
  density.}
\label{fig:t-m}
\end{minipage}
\end{figure} 


In the framework of the statistical model, one obtains a relation between the temperature and the baryonic chemical potential by fixing the value of the $K^-/K^+$ ratio. The l.h.s of 
Fig.~\ref{fig:t-m} shows the values of temperature and chemical
potential compatible with a value of the inverse ratio $K^+/K^-$ of around 30 
for the approaches
discussed above. The dot-dashed line stands for a free gas, similar to
the calculations of Ref.~\cite{Cleymans}. The on-shell approach (dotted
line) does not represent the broad-band effect, as already mentioned.
But due to the enhanced attraction felt by the $K^-$ mesons for higher
densities, the chemical potential $\mu_B$ compatible with the value of
the experimental ratio measured also increases for a given temperature.
When the $K^-$ spectral density containing s-wave components is used
(dashed line), two possible solutions that are compatible with the ratio
emerge. Finally, a band of chemical potentials $\mu_B$ up to 850 MeV at
a temperature of $T \approx 35$ MeV appears, when both, s- and p-wave
contributions are considered (solid line). However, in the latter case,
the temperature is too low to be compatible with the measured temperature
and the corresponding freeze-out densities
are too small (up to $0.02 \rho_0$ only), so we can hardly speak of a ``broad
band'' feature in the sense of that of Brown et al.  In the r.h.s.\ of
Fig.~\ref{fig:t-m} we represent the temperature and chemical potential
for different values of the ratio when the full $K^-$ spectral density
is used. We observe that the ratio is substantially lower at the more plausible temperature T $\approx$ 70 MeV, being more likely around 15 for a large region of baryonic chemical potential.  Note that
this reduced ratio translates into an overall enhanced production of
$K^-$ by a factor of 2 compared to the experimental value. This effect
is a consequence of the additional strength of the $K^-$ spectral
density at low energies. The Boltzman factor amplifies the contribution of the low energy region of the spectral function so that this additional strength is becoming the main reason for the overall enhanced production of the $K^-$ in the medium.

\section{Conclusions} 

The influence of a hot and dense medium on the properties of the hadrons
involved in the determination of the $K^-$/$K^+$ ratio has been studied. We have focused our attention on incorporating the effects of the antikaon properties in the medium at finite temperature.

It is found that the
temperature and chemical potential compatible with a given ratio depend
very strongly on the approach used for the in-medium properties of the
$K^-$ meson ($T \approx 35$ MeV and $\mu_B$ up to $850$ MeV for $K^+$/$K^-=30$
with the full $K^-$ spectral density).

The ``broad-band equilibration'' advocated by Brown, Rho and Song is
not achieved in the on-shell approach.  This behaviour is only
observed when $K^-$ is described by the full spectral density due to
the coupling of the $K^-$ meson to $YN^{-1}$ states.  However, the
$K^-/K^+$ ratio is in excess by a factor of 2 with respect to the
experimental one. What needs to be clarified is how the particles get on-shell at the freeze-out and this has to be addressed in dynamical non-equilibrium effects. However, this study is left for forthcoming work.

\vspace{-0.5cm}
\section*{Acknowledgments}

This work is partially supported by DGICYT project BFM2001-01868, by
the Generalitat de Catalunya project 2001SGR00064 and by NSF grant
PHY-03-11859. L.T. also wishes to acknowledge support from the
Alexander von Humboldt-Foundation.


\begin{thebibliography}{999}
  
\bibitem{Kaplan} D. B. Kaplan and A. E. Nelson, Phys. Lett.  {\bf B 175},
 57 (1986);
 {\it ibid.} {\bf B 179}, 409(E) (1986). 

\bibitem{Oeschler} H. Oeschler, J. Phys. {\bf G 28}, 1787 (2002); P.
  Senger, Acta Phys. Polon. {\bf B 31}, 2313 (2000); P. Senger, Nucl.
  Phys. {\bf A 685}, 312 (2001); C. Sturm et al., J. Phys. {\bf G 28},
  1895 (2002).
  
\bibitem{Barth} R. Barth et al., Phys. Rev. Lett. {\bf 78}, 4007
  (1997); F. Laue et al., Phys. Rev. Lett. {\bf 82}, 1640 (1999).
  
\bibitem{Menzel} M.  Menzel et al., Phys. Lett. {\bf B 495}, 26
  (2000).

\bibitem{Juergi} J. Schaffner-Bielich, V. Koch, and M. Effenberger,
  Nucl.  Phys. {\bf A 669}, 153 (2000).

  
\bibitem{Forster} A. F\"orster, PhD Thesis, TU Darmstadt (2002).
  
\bibitem{Hartnack} Ch. Hartnack, H. Oeschler, and J. Aichelin, Phys.
  Rev. Lett. {\bf 90}, 102302 (2003).
  
\bibitem{Cleymans} J. Cleymans, D. Elliot, A. Ker\"anen, and E.
  Suhonen, Phys.  Rev. {\bf C 57}, 3319 (1998); J. Cleymans, H.
  Oeschler, and K. Redlich, Phys. Rev.  {\bf C 59}, 1663 (1999); J.
  Cleymans, and K. Redlich, Phys. Rev. {\bf C 60}, 054908 (1999).
  
\bibitem{Brown} G. E. Brown, M. Rho and C. Song, Nucl. Phys. {\bf A
    690}, 184c (2001); G. E. Brown, M. Rho and C. Song, Nucl. Phys.
  {\bf A 698}, 483c (2002)
  
\bibitem{Laura02} L. Tol\'os, A. Ramos, and A.  Polls, Phys. Rev. {\bf
    C 65}, 054907 (2002).
  
\bibitem{Laura03} L. Tol\'os, A. Polls, A. Ramos, and J.
  Schaffner-Bielich, Phys. Rev. {\bf C 68}, 024903 (2003).
  
\bibitem{Balberg} S. Balberg, and A. Gal, Nucl. Phys. {\bf A 625}, 435
  (1997).
  
\bibitem{Mares} J. Mares, E. Friedman, A. Gal, and B. K. Jennings,
  Nucl.  Phys. {\bf A 594}, 311 (1995); J. Dabrowski, Phys. Rev. {\bf
    C 60}, 025205 (1999).
  
\bibitem{Laura01} L. Tol\'os, A. Ramos, A. Polls, and T. T. S. Kuo,
  Nucl. Phys. {\bf A 690}, 547 (2001).
  
\end{thebibliography}
\end{document}